\documentclass[12pt]{article}
\usepackage {amsmath}
\usepackage[dvips]{graphicx}
\usepackage{feynmp}
\usepackage{amssymb}
\usepackage{cite}
\newcommand{\be}{\begin{equation}}
\newcommand{\ee}{\end{equation}}
\newcommand{\bear}{\begin{eqnarray}}
\newcommand{\eear}{\end{eqnarray}}

\newcommand{\lapproxeq}{\lower .7ex\hbox{$\;\stackrel{\textstyle  
<}{\sim}\;$}} 
\newcommand{\gapproxeq}{\lower .7ex\hbox{$\;\stackrel{\textstyle  
>}{\sim}\;$}} 
\newcommand{\stackdown}[2]{\lower 1.4ex\hbox{$\;\stackrel{\textstyle{#1}}  
{\scriptstyle{#2}}\;$}}
\newcommand{\beq}{\begin{equation}} 
\newcommand{\eeq}{\end{equation}} 
\newcommand{\ba}{\begin{eqnarray}}
\newcommand{\ea}{\end{eqnarray}}

\newcommand{\bea}{\begin{eqnarray}}
\newcommand{\eea}{\end{eqnarray}}

\makeatletter 
\def\slash{\@ifnextchar[{\fmsl@sh}{\fmsl@sh[0mu]}} 
\def\fmsl@sh[#1]#2{% 
  \mathchoice 
    {\@fmsl@sh\displaystyle{#1}{#2}}% 
    {\@fmsl@sh\textstyle{#1}{#2}}% 
    {\@fmsl@sh\scriptstyle{#1}{#2}}% 
    {\@fmsl@sh\scriptscriptstyle{#1}{#2}}} 
\def\@fmsl@sh#1#2#3{\m@th\ooalign{$\hfil#1\mkern#2/\hfil$\crcr$#1#3$}} 
\makeatother 
\begin{document}
\begin{titlepage}  
%%%%%%%%%%% 
\begin{flushright} 
\parbox{4.6cm}{UA-NPPS/BSM-04/09 }
\end{flushright} 
%%%%%%%%%% 
\vspace*{5mm} 
\begin{center} 
{\large{\textbf { Five-Dimensional Black Hole  String Backgrounds and Brane Universe Acceleration.
}}}\\
\vspace{14mm} 
{\bf C. ~\ Chiou-Lahanas}, \,{\bf G. A.~\ Diamandis} and {\bf B. C.~\ Georgalas} 
{\footnote{email: \, cchiou@phys.uoa.gr, \, gdiam@phys.uoa.gr, \, vgeorgal@phys.uoa.gr}}
\vspace*{6mm} \\
  {\it University of Athens, Physics Department,  
Nuclear and Particle Physics Section,\\  
GR--15771  Athens, Greece}

\end{center} 
\vspace*{25mm} 
%%%%%%%%%%%%%%%%%%%%%%%%% 
\begin{abstract}
We examine background solutions of black hole type arising from the string effective action in five dimensions. We derive
the unique metric - dilaton vacuum which is a Schwarzschild type black hole. It is found that it can be extended to incorporate electric charge without changing the topology of the three space.
Kalb-Ramond charge may also be introduced if the three space is closed. The basic features of the cosmology induced on a three brane evolving in this background are also discussed. 
\end{abstract} 
\end{titlepage} 
\newpage 
\baselineskip=18pt 
%%%%%%%%%%%%%%%%%%%%%%%%%% Paper body %%%%%%%%%%%%%%%%%%%%%%%%%
%%%%%%%%%%%%%%%%%%%%%%%%%%%%%%%%%%%%%%%%%%%%%%%%%%%%%%%%%%%%%%%
%%%%%%%%%%%%%%%%%%%%%%%%%%%%
%%\newpage
\section*{Introduction}

The idea that the visible world may be realized as a brane in a higher dimensional space, where matter is confined is not new \cite{Rubakov}.  A  natural framework for this idea is String Theory in which 
higher dimensional spacetime arises by fundamental requirements.  The recognition that not all extra dimensions are of the order of Planck scale \cite{Antoniadis} and the discovery of D-branes \cite{Polchinski} naturally 
existing in String theory revived this idea in the recent years. The brane world scenaria \cite{review} have been extensively studied in order to address  fundamental questions such as the 
hierarchy problem \cite{Antoniadis1 , Randall, Gogbe}, the cosmological constant problem and issues regarding the early time cosmology.  In a large class of these models a Friedman-Robertson-Walker universe arises by the evolution of a three-dimensional brane in an AdS  or in an AdS black-hole spacetime \cite{cosmo, cosmo1, cosmo2, Chamblin, cosmo3, cosmo4, cosmo5, cosmo6, cosmo7, Kraus, Barcelo, Kehagias, Cai}. 
Of course the origin of the bulk Lagrangian and its relation to an underline fundamental theory is always under investigation. Towards this direction  higher dimensional black hole solutions have been extensively studied in various cases and especially in the context of Supergravity 
and String Theory \cite{Gibbons, GHS, Duff, Yazad, Aliev, Buchel, Castro}.
 
In this work we try to find exact background solutions for the tree level string action in five dimensional spacetime including a $U(1)$ field that may arise upon 
a partial compactification of a sixth dimension. 
In the first section we write the field equations and adopting 
Coulomb gauge both for the electromagnetic and the Kalb-Ramond fields and a static dilaton background we conclude with an 
ordinary system of equations depending only on the fifth dimension.
 For the metric we take the general form with maximally symmetric three-space. In the next 
section we proceed in solving the equations. As a first step we find the unique metric-dilaton background existing in the case of non-vanishing Liouville potential. 
This describes a Schwarzschild type black hole. This solution may be extended to include electric and Kalb-Ramond charges keeping the same form for the dilaton field. 
In the latter case the Kalb-Ramond field is non-trivial 
 only  for closed, $k=1$, three dimensional space. These backgrounds have the form of charged five-dimensional black holes. Examining the characteristics of the black holes we find 
 the mass and the temperature. In particular for the Schwarzschild type black hole the temperature is mass independent, a feature appearing in  black hole solutions arising from the String effective action \cite{Witten, chanmass}. This has as a consequence that in the charged case while the temperature increases with the mass, a feature shared by the AdS five-dimensional black holes \cite{Hawking},  there is a limiting 
temperature as $M \rightarrow  \infty$. 
In the last section  we briefly discuss the interesting features of the cosmology induced on a three brane to be identified with the observable universe. 
%%%%%%%%%%%%%%%%%%%%%%%%%%%%%%%%%%%%
\section*{Five-Dimensional String Effective Action Background}
The bosonic part of the low energy limit of the string  effective action is
\be
S\,=\, \frac{1}{2\kappa _0^2} \int d^Dx \sqrt{-G} e^{-2 \tilde{\phi} } \left \{ R\,+\,4(\nabla \tilde{\phi} )^2\,-\,
\frac{1}{12}H_{\mu \nu \rho }H^{\mu \nu \rho } \,-\, \frac{1}{4} F_{\mu \nu }F^{\mu \nu } \,-\,\Lambda  \right \}
\label{sfaction}
\ee
where $\Lambda$ is related to the central charge. In particular $\Lambda = \frac{2(D-26)}{3\alpha '}$ for the bosonic string, and $\Lambda = \frac{(\frac{3D}{2}-15)}{3\alpha '}$ for the superstring, 
  $\tilde{\phi} $ is the dilaton field, 
 $H_{\mu \nu \rho }\,=\, \nabla _{\mu } B_{\nu \rho }+ \nabla _{\rho  } B_{\mu \nu  } + \nabla _{\nu } B_{\rho \mu  }$ is the field strength of the antisymmetric tensor field $B_{\mu \nu }$. We 
also assume the existence of $U(1)$ gauge field  $A_{\mu}$ with $F_{\mu \nu }=\partial _\mu A_\nu -\partial _\nu A_\mu $ its field strength. 
Such a gauge field can be thought as a remnant of a partial compactification.\\
%%%%%%%%%%%
The curvature term can be cast to the Einstein-Hilbert form by the Weyl rescaling of the metric
$$
g_{\mu \nu }\,=\, e^{-\frac{2\phi }{D-2}} G_{\mu \nu }\, \quad \phi \, = \, \tilde{\phi} - \phi_0.
$$ 
After this rescaling the dilaton appears as a matter field with a Liouville-type potential and of particular couplings with the other fields
\be
S\,=\, \frac{1}{2\kappa ^2} \int d^Dx \sqrt{-g}  \left \{ R\,-\,\frac{4}{D-2}(\nabla \phi )^2\,-\,
\frac{1}{12}e^{-\frac{8\phi }{D-2}}H^2 \,-\, \frac{1}{4}e^{\frac{-4\phi }{D-2}} F^2 \,-\, V(\phi ) 
 \right \}
\label{efaction}
\ee
where $V(\phi )\,=\,\Lambda  e^{\frac{4\phi }{D-2}}$. Note that $\kappa ^2 =  e^{2 \phi_0} \kappa_0^2 = (8 \pi G_D)$ is related to the D-dimensional Newton's constant.
%%%%%%%%%%%%%%%%%%%%%%%%%%%%%%%%%%%%%%%%%%%%%%%%%%%%%%%%%%%%

For a five-dimensional space the field equations read
\bea
R_{\mu \nu }-\frac{1}{2}g_{\mu \nu }R &=& \sigma _1 \left( \nabla _{\mu } \phi \nabla _{\nu }
\phi - \frac{1}{2}g_{\mu \nu } (\nabla \phi)^2 \right) - \frac{1}{4} e^{\sigma _2 \phi } 
\left ( H_{\mu \rho \sigma }H_{\nu }^{\rho \sigma } - \frac{1}{6}g_{\mu \nu}H^2 \right ) \nonumber \\
 &+& \frac{1}{2}e^{\sigma _4\phi } \left ( F_{\mu \rho } F_{\nu }^{\rho } - \frac{1}{4} g_{\mu \nu } F^2 \right ) - \frac{\Lambda }{2} e^{\sigma _1\phi }g_{\mu \nu } \nonumber
\\
2 \Box \phi - \Lambda e^{\sigma _1 \phi }&+& \frac{1}{6}e^{\sigma _2 \phi }H^2 + \frac{1}{4}e^{\sigma _4 \phi }F^2 = 0  \nonumber \\
\nabla _ \rho  \left \{ e^{ \sigma _2 \phi } H ^\rho _{\mu \nu } \right \} &=& 0 \nonumber \\
\nabla _ \mu  \left \{ e^ {\sigma _4 \phi } F^{\mu \nu } \right \} &=& 0 
\eea
where $\sigma _1=-\sigma _4=4/3\, \quad \sigma _2= -2 \sigma_1 = -8/3$. \\
If we set
\be
H^{\mu \nu \rho } = \frac{1}{2} e^{- \sigma _2 \phi } E^{\mu \nu \rho \sigma \tau }
\mathcal{B}_{\sigma \tau }\,,
\label{dual}
\ee
where $E^{\mu \nu \rho \sigma \tau }$ is the totally antisymmetric tensor in five dimensions and
$ \mathcal{B}_ {\sigma \tau } = \partial _{\sigma } \mathcal{K}_\tau - \partial _{\tau  } \mathcal{K}_ \sigma $ is the field strength of a vector field $\mathcal{K}_{\mu }$, the equations become
\bea
R_{\mu \nu }-\frac{1}{2}g_{\mu \nu }R &=& \sigma _1 \left( \nabla _{\mu } \phi \nabla _{\nu } \phi - 
\frac{1}{2}g_{\mu \nu } (\nabla \phi)^2 \right) + \frac{1}{2} e^{-\sigma _2 \phi }
 \left ( \mathcal{B}_{\mu \rho  } \mathcal{B}_{\nu }^{\rho } - \frac{1}{4}g_{\mu \nu } \mathcal{B}^2 \right ) \nonumber \\
 &+& \frac{1}{2}e^{\sigma _4\phi } \left ( F_{\mu \rho } F_{\nu }^{\rho } - \frac{1}{4} g_{\mu \nu } F^2 \right ) - 
\frac{\Lambda }{2} e^{\sigma _1\phi }g_{\mu \nu } \nonumber 
\\
2 \Box \phi - \Lambda e^{\sigma _1 \phi }&-& \frac{1}{2}e^{-\sigma _2 \phi }\mathcal{B}^2 + 
\frac{1}{4}e^{\sigma _4 \phi }F^2 = 0 \nonumber  \\
\nabla _ {\mu  } \left \{ e^ {-\sigma _2 \phi } \mathcal{B} ^{\mu \nu } \right \} &=& 0 \nonumber  \\
\nabla _ \mu  \left \{ e^ {\sigma _4 \phi } F^{\mu \nu } \right \} &=& 0 \,.
\label{dualeq}
\eea
Note that the substitution (\ref{dual}) interchanges the equation of motion with the Bianchi identity for 
the Kalb-Ramond field and the equations 
(\ref{dualeq}) are those derived from the action
\be
S\,=\, \frac{1}{2\kappa ^2} \int d^5x \sqrt{-g}  \left \{ R\,-\,\frac{4}{3}(\nabla \phi )^2\,-\,
\frac{1}{4}e^{-\sigma _2 \phi } \mathcal{B}^2 \,-\, 
\frac{1}{4}e^{\sigma _4} F^2 \,-\, V(\phi )  \right \}\,.
\label{dualaction}
\ee

The general spherically symmetric five-dimensional metric with maximally symmetric three space reads:
\be
ds^2 \,=\, -C(y',t)dt^2 \,+\, D(y',t)dy'^2 \,+\, 2 E(y',t)dy'dt \,+\,y'^2F(y',t)d\Omega _3^2
\label{metrican}
\ee
where $d\Omega _3^2 \,=\, \frac{dr^2}{1-kr^2}+r^2(d\theta ^2+sin^2\theta d\varphi ^2)$. \\
The choice of a maximally symmetric three space is made since it is appropriate for cosmological 
studies. Defining $y=y'F^{1/2}$ and following the well known diagonalization procedure,  the metric can always be brought in the standard form
\be
ds^2 \,=\, -\frac{V^2(y, t)}{A(y, t)}dt^2 \,+\, A(y,t)dy^2 \,+ \,y^2d\Omega _3^2\,.
\label{genmet}
\ee
Choosing  Coulomb gauge for both the electromagnetic potential $A_\mu (y, t) = (a(y,t), 0, 0, 0, 0)$ 
and the Kalb-Ramond field $\mathcal{K}_\mu (y, t) = (b(y , t),0,0,0,0)$,  field equations are easily integrated yielding:
\bea
 F_{yt} \,&=&\, a'(y, t) \,=\, \frac{e^{-\sigma _4 \phi (y, t)}V(y, t)}{y^3}\,q \nonumber \\
 \mathcal{B}_{yt} \,&=&\, b'(y, t) \,=\, \frac{e^{\sigma _2 \phi (y, t)}V(y, t)}{y^3} \,q_b \,,
\label{coulomb}
\eea
where prime denotes differentiation with respect to $y$ and $q\,, \,\,\,q_b$ are constants of 
integration related to the corresponding charges. This gauge choice and consequently  (\ref{coulomb}) 
gives electric type solutions for both vector fields.

Seeking for  static backgrounds, one can remark that demanding the dilaton to be static is adequate to enforce all fields to be static.  
Indeed looking at the simplest  of the 
metric equations (this with $\mu =1\,,\nu =2$) which has the form
\be
\frac{3 \dot{A}(y, t)}{2yA(y, t)} - \sigma _1 \dot{\phi }(y, t) \phi '(y, t) \,=\,0
\label{12}
\ee
we see that $\dot{\phi }(y, t)=0$ implies that $\dot{A}(y, t)=0$.  Then a linear combination of the ($\mu =\nu =1$) 
and ($\mu =\nu =2$) metric equations gives 
\be
\frac{4}{9} y \phi '^2(y) - \frac{V'(y, t)}{V(y, t)} = 0\,,
\label{vsol}
\ee
yielding
\[
V(y, t) = f(t) e^{\int^y \frac{4}{9}z\phi '^2(z)dz}\,.
\]
The arbitrary function $f(t)$ can be absorbed into a time redefinition leaving $V$ static and thus 
making the metric static. The fields $a'$ and $b'$
become in turn static also. 

The system of the remaining independent equations for the static case becomes
\bea
\frac{3U'}{2y} + \left(\frac{3}{y^2}+\frac{2}{3}\phi '^2  \right)U + \frac{\Lambda }{2}e^{4\phi /3} - 
\frac{3k }{y^2} + \frac{e^{-8\phi /3}}{4y^6}q_b^2 + \frac{e^{4\phi /3}}{4y^6}q^2 \,&=&\,0  \nonumber \\
\frac{U''}{2} + \left(\frac{2}{y}+ \frac{3V'}{2V}  \right)U'+ \left(\frac{1}{y^2}+\frac{2}{3}\phi '^2 
+ \frac{2V'}{yV} + \frac{V''}{V} \right)U &+& \nonumber \\
 \frac{\Lambda }{2}e^{4\phi /3} - 
\frac{k }{y^2} - \frac{e^{-8\phi /3}}{4y^6}q_b^2 -\frac{e^{4\phi /3}}{4y^6}q^2 \,&=&\,0 \nonumber \\
\phi ' U' + \left(\frac{3\phi '}{y}+\frac{V'}{V}\phi ' + \phi ''  \right)U - \frac{\Lambda }{2}e^{4\phi /3} + 
\frac{e^{-8\phi /3}}{4y^6}q_b^2 - \frac{e^{4\phi /3}}{4y^6}q^2 \,&=&\,0 \,,
\label{fullstsys}
\eea
where $U=A^{-1}$.
Note that $V'/V$ is solved in terms of the dilaton through (\ref{vsol}) and as a consequence a non-trivial dilaton 
implies asymmetric form of the five-dimensional metric.
 All the quantities are taken to be dimensionless. The right dimensionality
can always be reinstablished by appropriate powers of $\alpha ' $ and $G_5$.\\
    
%%%%%%%%%%%%%%%%%%%%%%%%%%%%%%%%%%%%%%%%%%%%%%%%%%%%%%%%
\section*{Black Hole Solutions}
In order to find static solutions we explore first the case of a "metric-dilaton string background" that is we 
take $q=q_b=0$.

In this case we remark that linear combinations of the three equations lead to a system which does not depend on the 
parameter  $\Lambda $ but only on $k$.  The system can be cast in the form
%%%%%%%%%%%%%%%%%%%%%%%%%%%%%%%%%%%%%%%%%%%%%%%%%%%%%%%%%%%%%
\bea
\left( \frac{1}{2 \phi '}-\frac{y}{3}  \right )U'+
 \left[ \left( \frac{2}{y}+\frac{V'}{V} \right ) \left( \frac{1}{2 \phi '} - \frac{y}{3}  \right )-
\frac{y}{3}\left( \frac{1}{y}+\frac{\phi ''}{\phi '}  \right ) \right]U \,& = &\,\frac{k}{y \phi '}, \nonumber \\
\left( \frac{d}{dy}+\frac{3}{y} +\frac{V'}{V} \right ) \left( \frac{d}{dy} +\frac{2V'}{V} - \frac{2}{y} \right )U \,&=&\, -\frac{4k}{y^2}, \nonumber \\
\left( \frac{d}{dy}+\frac{3}{y} +\frac{V'}{V} \right ) \left( \frac{d}{dy} +\frac{2V'}{V} + \frac{4}{3} \phi ' \right )U \,&=&\, 0.
\label{homog}
\eea
%%%%%%%%%%%%%%%%%%%%%%%%%%%%%%%%%%%%%%%5
From the above equations the homogeneous one admits the general solution
\be
U \,=\, (A + B u) \frac{e^{-\frac{4 \phi}{3}}}{V^2} \,, \quad u' = \frac{e^{\frac{4 \phi}{3}} V}{y^3}
\ee
Using this expression for $U$ it can be proven that in any case the dilaton is enforced to 
satisfy the simple equation
\[  
y \phi '' + \phi ' \,=\, 0 \,.
\]
Inserting  the general solution $\phi (y) \,=\, m + b ln y$ of this equation into the 
system (\ref{fullstsys}) we find  the unique result 
\be
k\,=\,0\,, \quad \phi (y) \,=\, m - \frac{3}{2} ln y \,, \quad V(y)\,=\,y
\label{vacsol}
\ee
and for the constants we get $A=-\frac{\Lambda }{9}e^{\frac{8m}{3}}$ while $B$ remains arbitrary. Note that in our 
case $\Lambda $ is taken to be  nonzero and since we work in subcritical dimensions  it is negative. So we obtain the vacuum solution
\bea 
U(y) \, &=& \, -\frac{1}{9} \Lambda e^{4m/3} - \frac{M}{y^3}  \,, \nonumber \\
V(y) \, &=& \, y                 \,, \nonumber \\
\phi (y) \, &=& \,   m - \frac{3}{2} ln y  \,. 
\label{vacsolution}
\eea
where $M=B/3$. As we will discuss in the following the above solution describes a five-dimensional black hole of Schwarzschild  type.
Note that the logarithmic dependence  found for the dilaton field is the one demanded so that the homogeneous part of the full system (\ref{fullstsys}) has non-trivial solution. 

Keeping this form of the dilaton field  we see that the following solution of the full system with non-trivial vector fields exists:
\bea 
U(y) \, &=& \, \frac{1}{9} \left(-\Lambda + \frac{q^2}{2y^6}  \right) e^{4m/3} - \frac{M}{y^3} + \frac{4k}{9} \,, \nonumber \\
V(y) \, &=& \, y                 \,, \nonumber \\
\phi (y) \, &=& \,   m - \frac{3}{2} ln y  \,, \nonumber \\
a'(y)     \, &=& \, \frac{q}{y^4} e^{4m/3}           \,, \nonumber \\
b'(y)     \, &=& \,  q_b y^2 e^{-8m/3}\,, \,\,  \quad   q_b^2 \,=\, 4 k e^{8m/3}\,, 
\label{solution}
\eea
 describing a five-dimensional electrically charged black hole solution with non-trivial Kalb-Ramond   charge which, 
unlike the electric charge, may be present 
only for $k =1$.  

Let us note that the string case with non-trivial dilaton potential escapes from the black hole backgrounds already found in 
Maxwell-Dilaton gravity \cite{chanmass, Sheykhi}.
The novel feature here is the proof, analogous to Birhkoff theorem, 
of the uniqueness of the metric-dilaton background adopting the general form of the five-dimensional metric (\ref{genmet})
without any further  ansatz\footnote{The uniqueness of the black hole solutions in five-dimensional Dilaton gravity is addressed also in \cite{Kiritsis}.}. Moreover the use of the more general form of the metric  allows to find
the solution with non-trivial Kalb-Ramond field for the String Effective action  \cite{DeRisi}. 
%%%%%%%%%%%%%%%%%%%%%%%%%%%%%%%%%%%%%%%%%%%%%%%%%%%%%%%
\subsection*{Characteristics of the Black Hole Solutions.} 

The line element for the solution found above is of the form:
\be
ds^2 \,=\, -y^2U(y)dt^2 + \frac{dy^2}{U(y)} + y^2d\Omega _3^2
\label{line}
\ee
where 
\[
U(y) = \lambda - \frac{M}{y^3} + \frac{Q^2}{y^6}\, \quad \lambda = \frac{4 k }{9}- \frac{\Lambda e^{4m/3}}{9}\,,
 \quad Q^2= \frac{q^2 e^{4m/3}}{18}\,.
\]
The maximally symmetric 3-space is described by:
\be
d \Omega ^2_3 = \Big \{ 
\begin{array}{c} d \chi ^2+sin^2\chi (d \theta ^2 + sin^2 \theta d \varphi ^2 )\,, \quad \,\, k \,=\, 1 \\
dr ^2+r^2  (d \theta ^2 + sin^2 \theta d \varphi ^2) \,, \quad   \,\,  k\,=\, 0  
\end{array}
\ee
This line element exhibits a physical, curvature, singularity at $y=0$ where $\phi \rightarrow + \infty$, meaning divergence of 
the string coupling constant at the singularity.
When $Q \neq 0$ we have two horizons located at
\[
0 < y_\pm ^3 = \frac{M\pm  \sqrt{M^2-4\lambda Q^2}}{2\lambda }
\]
provided that $M > 2 \mid Q\mid  \sqrt{\lambda } $,  with $y_+$ denoting the event horizon. The  case where 
$M = 2 \mid Q\mid  \sqrt{\lambda } $, gives the extremal black hole. If  $Q = 0$ we have a Schwarzschild-type
black hole.

The metric has "unusual" asymptotics in the sense that it is not asymptotically flat nor Anti-de-Sitter. In order to determine the mass
we use the quasi-local mass  formalism described in \cite{Brown,  Poletti, chanmass, Hawking1, surmass}. The  quasi-local mass in this case
 is defined  as:
\be
m(y_b) \, = \, \frac{y \sqrt{U(y_b)}}{8 \pi} \left [ \lambda ^{1/2} - \sqrt{U(y_b)} \right ] \left[ 
\frac{d}{dy} S_3 (y)  \right ] _{y=y_b} 
\label{quasimass} 
\ee
where $S_3(y)=2 \pi ^2 y^3$ is the surface of the four-dimensional sphere with radius $y$. For the metric in (\ref{line}) 
we get:
\be
m(y_b) = \frac{3 \pi}{4} \frac{ \sqrt{\left ( \lambda  - \frac{M}{y_b ^3} + \frac{Q^2}{y_b^6} \right )}
  \left(   \frac{M}{y_b ^3} - \frac{Q^2}{y_b^6}  \right ) y_b^3 }
{ \sqrt{\lambda } + \sqrt{\lambda - \frac{M}{y_b ^3} + \frac{Q^2}{y_b^6} }} \,.
\label{linequasi}
\ee
Finally in the limit $y_b \rightarrow \infty $ the black hole mass is found to be:
\be
m_\infty = \frac{3 \pi M}{8} \,.
\label{mass}
\ee
Note that the mass parameter $M$ in (\ref{line}) is related to the black hole mass by the known relation 
$M= \frac{8m_\infty }{3 \pi}$ for the mass of the  Reissner - Nordstrom black hole in five dimensions \cite{Das, Mora}.
%%%%%%%%%%

As far as the temperature is concerned we have to avoid the conical singularity of the metric at the horizon. This is achieved by 
defining new coordinates $(t, \rho) $ in the $(t, y)$ subspace. The new variable $\rho $ is defined as 
\[
\rho = \int_{y^+}^y \frac{du}{U^{1/2}(u)}\,,
\]
where $y_+$ is the event horizon. In terms of these coodinates the metric becomes
\[
ds^2 = -G(\rho ) \rho ^2 dt^2 + d\rho ^2 + y^2(\rho )d\Omega _3^2 \,,
\]
where $G(\rho) \rho ^2 = y^2 U(y)$. \\

The variable $\rho $ comes to be:
\be
\sqrt{\lambda}  \rho (y) \,=\, y \sqrt{ \left( 1 - \frac{y_+^3}{y^3} \right) \left( 1 - \frac{y_-^3}{y^3}  \right)}  + A(y; y_+, y_-) - A(y_+; y_+, y_ -)
\label{rho}
\ee
where 
\[  
A(y; y_+, y_-) \,=\, 
\frac{y_+^3 + y_-^3}{4y^2} F_1 (\frac{3}{2} , \frac{1}{2}, \frac{1}{2}, \frac{5}{3}, \frac{y_+^3}{y^3}, \frac{y_-^3}{y^3}) - 
\frac{2 y_+^3  y_-^3}{5y^5} F_1 (\frac{5}{3} , \frac{1}{2}, \frac{1}{2}, \frac{8}{3}, \frac{y_+^3}{y^3}, \frac{y_-^3}{y^3})
\]
$F_1$ is the Appell Hypergeometric function in two variables. The constant $A(y_+; y_+, y_ -)$, takes the simpler form
\[  
A(y_+; y_+, y_ -) \,=\, \frac{\sqrt{\pi} (y_+^3 + y_-^3)}{4y_+^2} \, \frac{\Gamma (\frac{5}{3})}{\Gamma (\frac{7}{6})}\, _2F_1 
( \frac{2}{3}, \frac{1}{2}, \frac{7}{6}, \frac{y_-^3}{y_+^3}) \,-\, 
\frac{\sqrt{\pi} (2y_-^3)}{5y_+^2} \, \frac{\Gamma (\frac{8}{3})}{\Gamma (\frac{13}{6})}\, 
_2F_1 
( \frac{5}{3}, \frac{1}{2}, \frac{13}{6}, \frac{y_-^3}{y_+^3}).
\]
Note that for $y \gg y_+$, 
 the line element becomes 
\[
ds^2 \approx -\lambda y^2 dt^2 + \frac{1}{\lambda } dy^2 +y^2 d\Omega _3 ^2 \,.
\]
This means that the space becomes asymptotically Rindler explaining the fact that the mass parameter found previously 
coincides with that of an asymptotically flat Reissner-Nordstrom black hole in five dimensions. \\
Now for the temperature we need the limit of $G^{1/2}(\rho (y))$ as $y$ approaches the horizon $y_+$. This limit turns out to be 
\[
G^{1/2} (0) \, = \, \frac{1}{2} \sqrt{ \left [ y^2 U(y) \right ]'_{y=y_+} U'(y_+)} \,=\, \frac{1}{2} y_+ U'(y_+)\,.
\]
The periodicity in the Euclidean time $\tau = -it$ gives the temperature $T_H$ through the relation \cite{chanmass, Bousso, Horowitz, Mora}
\be
T_H \,=\, \frac{1}{4 \pi} G^{1/2}(0) \,=\, \frac{3 \lambda }{8 \pi} \frac{y_+^3-y_-^3}{y_+^3} \,.
\label{temp}
\ee
Regarding the temperature let us point out the basic properties. As far as the Schwarzschild case is concerned we see that the 
temperature is mass independent. This feature although peculiar appears for certain solutions of the string case in 
two and four dimensions \cite{Witten, chanmass}. 
For the charged case
 the temperature increases with the mass
 exhibiting  a maximum value $T_{max} =  \frac{3 \lambda }{8 \pi}$ as $M \rightarrow \infty $. 
%%%%%%%%%%%%%%%%%%%%%%%%%%%%%%%%%%%%%%%%%%%%%%%%%%%%%%%%%%%%%%%%%%%%%%%%%%%%%%%%%%%%%%%%%%%%%%
%%%%%%%%%%%%%%%%%%%%%%%%%%%%%%%%%%%%%%%%%%%%%%%%%%%%%%%%%%%%%%555
\section*{Cosmological Implications}
In this section we will briefly discuss the characteristics of the induced cosmology on a brane evolving in the above 
background. Following the steps in \cite{Kraus, Barcelo, Brax, Kanti1, Tetradis} we introduce a brane at the point $y \,=\, R$. 
The world volume spanned by the brane is determined by the vector $X^{\mu }\,=\,(t(\tau ),\, R(\tau ),\, x^i)$ where $\tau $ is the proper time. 
The five-dimensional line element is
\[
ds_5^2 \,=\, G_{\mu \nu } dx^\mu  dx^\nu \,=\, -A dt^2 \,+\, B dy^2 \,+\, y^2 \,d\Omega _3^2\,.
\]
The metric induced on the brane, $g_{ab}\,=\, \frac{\partial X^{\mu }}{\partial \xi  ^a} \, \frac{\partial X^\nu }{\partial \xi  ^b} \, G_{\mu \nu }$, where $\xi ^a \,=\, (\tau ,\,x^i)$ yields the line element
\[
ds_{4}^2 \,=\, -d\tau ^2 + R^2(\tau ) d\Omega _3^2\,,
\]
provided that $\frac{dt}{d\tau } \,=\, \sqrt{\frac{1+B \dot{R}^2}{A}}$. The velocity vector is 
$u^{\mu } \,=\, (\frac{dt}{d\tau },\, \dot{R},\, \vec{0})$ where $\dot{R}$ is the derivative with respect to the proper time $\tau $.
The unit vector normal to the brane directed inside the bulk space described by the original five-dimensional metric and uniquely determined by the conditions $u_\mu  \eta ^\mu \,=\,0$ and $\eta _\mu  \,\eta ^\mu \,=\,1 \,,$ is
$\eta ^{\mu } \,=\, \frac{1}{\sqrt{AB}}\,(-B\,\dot{R},\, -A\, \frac{dt}{d\tau },\, \vec{0}) $ .
Assuming that the brane describing the visible spacetime is the boundary separating two regions of the five-dimensional spacetime with two different metrics the Israel junction conditions impose that
\[
K_{ab }^+ - K_{ab }^- \,=\, - 8\pi G \left[ T_{ab} - \frac{1}{3}\,T\,g_{ab}\right]
\] 
where 
\[
K_{ab} \,=\, \frac{1}{2} \frac{\partial X^{\mu }}{\partial \xi  ^a} \, \frac{\partial X^\nu }{\partial \xi  ^b} \, (\eta _{\mu ; \nu } 
\,+\, \eta _{\nu ;\mu }) \]
  is the extrinsic curvature of the brane, $T_{ab}$ is the energy momentum tensor on the brane and $T$
its trace. Imposing a $Z_2$ symmetry we get
\[
K_{ab }\,=\, - 4\pi G \left[ T_{ab} - \frac{1}{3}\,T\,g_{ab}\right].
\] 
In our case the components of the extrinsic curvature are
\bea
K_{ij} &\,=\,& -  \frac{1}{R} \, \sqrt{U+ \dot{R}^2} \,  g_{ij} \nonumber \\
K_{\tau \tau } &\,=\,& \frac{1}{V} \, \frac{d}{dR} \,(V\, \sqrt{U + \dot{R}^2}) \,,
\eea
where $V\,=\, \sqrt{AB}$ and $U\,=\,B^{-1}$.

For our simple consideration in this work we will be restricted in an empty brane with just 
a tension term as the brane Lagrangian \cite{Kanti1}, that is $T_{ab} \,=\, -\sigma g_{ab}$. The $\{ij\}$ Israel junction condition, yield the Friedmann equation on the brane:
\be
H^2 \,=\, \left( \frac{\dot{R}}{R} \right)^2 \,=\, \left( \frac{4\pi G \sigma}{3} \right)^2 \,-\, \frac{U(R)}{R^2} \,=\, 
\left( \frac{4 \pi G \sigma}{3} \right)^2 \,-\, \frac{\lambda}{R^2}  + \frac{M}{R^5} - \frac{Q^2}{R^8}\,.
\label{shubble}
\ee
We remark that although we have an asymmetric form of the five-dimensional  metric, 
the Friedman equation coincides  with that  in \cite{Kraus} since $K_{ij}$ does not depend on $V$.

The basic remarks regarding the above equation are:
The 3-brane tension cannot be balanced with the "five-dimensional cosmological constant" which in our case is the dilaton potential.
This implies that if we insist in the tension  $\sigma$ we have always a minimum radius and an infinitely expanding universe. 
No cyclic solutions appear \cite{Kanti1}.
The acceleration always tends asymptotically to unity. 
The acceleration can easily be evaluated in terms of the scale factor $R$ through (\ref{shubble}):
\[
q \,=\, \frac{R \ddot{R}}{\dot{R}^2}\, =\, \frac{R}{2H^2}\frac{dH^2}{dR}\,+\,1\,.
\] 
Depending on the relations between the parameters we 
can achieve evolution where the accelerating period follows a decelerating one. The transition point lies outside the 
 event horizon. Moreover this transition occurs at a phenomenologically acceptable $z$ with respect to the present day acceleration (see Figure 1). This feature proposes that the above background may be useful for the study of late time cosmology.
 
There are two obstacles for such an interpretation. The first is the constant asymptotic value for the Hubble parameter and the unit asymptotic value for the acceleration. This is due to the case considered in this work with no matter on the brane. The inclusion of non-trivial energy momentum tensor on the brane will in general cure this unpleasant characteristic. The most serious drawback is that the $\{ \tau \tau  \}$  junction condition
\[
\frac{1}{V} \, \frac{d}{dR} \,(V\, \sqrt{U + \dot{R}^2}) \,=\, \frac{4\pi G \sigma }{3}\,,
\]
is not satisfied. This is a consequence of the asymmetric form of the solution of the bulk metric, $V\,\neq\,1$. Besides upon the inclusion of matter this condition is not compatible with the usual state equations. Such a situation is known to appear in this type of metrics \cite{Chamblin, Brax}. The ways out of this problem is either the consideration of backreaction effects rendering the bulk metric to be time-dependent or by inclusion of dilaton-dependent terms in the brane Lagrangian. One other possibility is to try to satisfy the junction condition keeping the state equation for usual matter and imposing some kind of exotic matter with unusual state equation. Such kind of matter comes out if non-critical evolution is assumed \cite{case}. All these possibilities as well as the relaxing of the $ Z_2$ symmetry need of course further study and work is in progress in these directions.

%%%%%%%%%%%%%%%%%%%%%
%\newpage
\begin{figure}
\begin{center}
%\vspace{0.5cm}
\includegraphics[width=12cm, height=5cm]{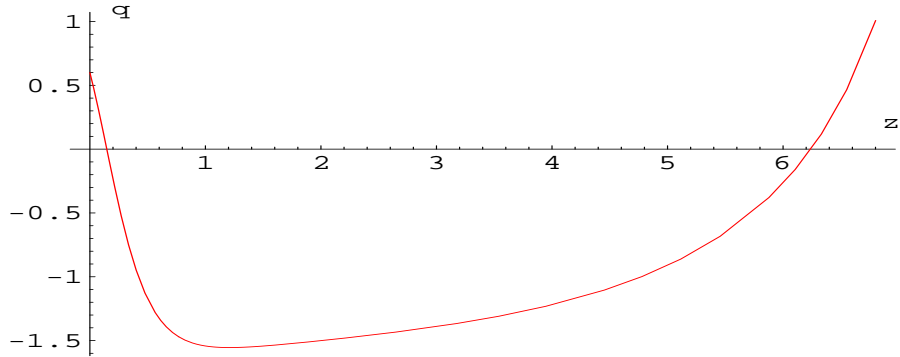}
\end{center}
\vspace{0.5cm}
\caption{ The acceleration as a function of $z$. We see the transition from the decelerating to the accelerating phase occuring at $z\simeq 0.2$ }
\label{Figure 1}  
\end{figure}
%%%%%%%%%%%%%%%%%%%%%%%%%%%%%%%%%%%%%%%

%\newpage
\section*{Summary and Discussion}

In this work we examine background solutions of the String Effective action in five dimensions. In the beginning we derive the 
unique dilaton - metric static vacuum with non-vanishing dilaton potential. This solution represents a Schwarzschild type black hole.  
It is also found that this solution can be extended to incorporate electric charge leading to charged five-dimensional black hole. Furthermore 
Kalb-Ramond charge may be also present. In this case the three-dimensional space has to be closed ($k=1$). 
The uniqueness statement is not valid for non-trivial vector fields. Nevertheless for   logarithmic dilaton field there are 
only two solutions of (\ref{fullstsys}). The one analyzed in this work and the following with vanishing dilaton potential $\Lambda  = 0$:
\bea 
U(y) \, &=& \,  \frac{q_b^2}{2y^{12}}  e^{-8m/3} - \frac{M}{y^6} + \frac{k}{9} \,, \nonumber \\
V(y) \, &=& \, y^4                 \,, \nonumber \\
\phi (y) \, &=& \,   m + 3 ln y  \,, \,\,  \quad   q^2 \,=\, 8 k e^{-4m/3}\,, \nonumber \\
a'(y)     \, &=& \, q y^5 e^{4m/3}           \,, \nonumber \\
b'(y)     \, &=& \, \frac{ q_b}{ y^2} e^{-8m/3}\,.  
\label{solution}
\eea
Note here the interchange between the electric and Kalb-Ramond charges. This background, although  interesting 
by itself, is not analyzed in this work since it is not 
connected to the unique metric-dilaton vacuum. A last comment regarding the background is that for $k=0$ we are obliged to have 
negative $\Lambda $ in order to have the black hole character. This is justified as is discussed in the first section. Nevertheless in the case 
with $k=1$, positive $\Lambda $ is also allowed coming closer to the case described in \cite{aben}.

The mass and the temperature of these black holes  have been calculated. The mass is found to agree with what is expected for asymptotically flat five dimensional black holes. For the  temperature we found that for the Schwarzschild type black hole the temperature is mass independent and for the charged one we have, for given charge, temperature increasing with the mass but with a limiting value.

The  cosmology induced on a three brane evolving in this background is also briefly discussed.  For this particular type of cosmological evolution we find that we always have a  minimum radius in accordance with other authors. In our case we do not have cyclic solutions, that is the radius goes always to infinity. The good feature is that for a wide range of the parameters involved we find accelerating phase following a decelerating one. The transition occurs in a phenomenologically  acceptable range. The basic problem of using such scenaria to study late time cosmology is pointed out.  A more thorough study of the thermodynamics of the black hole solutions found and their cosmological implications, mainly in the direction of solving the problem raised is in progress.     

%%%%%%%%%%%%%%%%%%%%%%%%%%%%%%%%%%%%%% 
\section*{Acknowledgments}
This work is co-funded by the Research Special Account of the National and Kapodestrian University of Athens. \\
%%%%%%%%%%%%%%%%%%%%%%%%%%%%%%%%

%%%%%%%%%%%%%%%%%%%%%%%%%%%%%%%%%%%%%%%%%%%%%

\end{document}